# High mobility in a van der Waals layered antiferromagnetic metal

Shiming Lei[1], Jingjing Lin[2], Yanyu Jia[2], Mason Gray[3], Andreas Topp[4], Gelareh Farahi[2], Sebastian Klemenz[1], Tong Gao[2], Fanny Rodolakis[5], Jessica L. McChesney[5], Christian R. Ast[4], Ali Yazdani[2], Kenneth S. Burch[3], Sanfeng Wu[2], N. Phuan Ong[2] and Leslie M. Schoop[1]*

[1]*Department of Chemistry, Princeton University, Princeton, New Jersey 08544, USA*

[2]*Department of Physics, Princeton University, Princeton, New Jersey 08544, USA*

[3]*Department of Physics, Boston College, Boston, MA 02467, USA*

[4]*Max-Planck-Institut für Festkörperforschung, Heisenbergstraße 1, D-70569 Stuttgart, Germany*

[5]*Argonne National Laboratory, 9700 South Cass Avenue, Argonne, Illinois 60439, USA*

Email: lschoop@princeton.edu

**Abstract**

Magnetic van der Waals (vdW) materials have been heavily pursued for fundamental physics as well as for device design. Despite the rapid advances, so far magnetic vdW materials are mainly insulating or semiconducting, and none of them possesses a high electronic mobility – a property that is rare in layered vdW materials in general. The realization of a magnetic high-mobility vdW material would open the possibility for novel magnetic twistronic or spintronic devices. Here we report very high carrier mobility in the layered vdW antiferromagnet $GdTe_3$. The electron mobility is beyond 60,000 $cm^2$ $V^{-1}$ $s^{-1}$, which is the highest among all known layered magnetic materials, to the best of our knowledge. Among all known vdW materials, the mobility of bulk $GdTe_3$ is comparable to that of black phosphorus, and is only surpassed by graphite. By mechanical exfoliation, we further demonstrate that $GdTe_3$ can be exfoliated to ultrathin flakes of three monolayers, and that the magnetic order and relatively high mobility is retained in ~20-nm-thin flakes.

VdW materials are the parent compounds of two-dimensional (2D) materials, which are currently actively studied for new device fabrications (*1*) involving the creation of heterostructure stacks (*2*) or twisted bilayers (*3*) of 2D building blocks. Magnetic vdW materials have recently led to the observation of intrinsic magnetic order in atomically thin layers (*4-12*), which was followed by exciting discoveries of giant tunneling magnetoresistance (*13-16*) and tunable magnetism (*17-19*) in such materials.

So far, the known magnetic vdW materials (ferro- or antiferromagnetic) that can be exfoliated are limited to a few examples, such as: $CrI_3$ (*4*), $Cr_2Ge_2Te_6$ (*5*), $FePS_3$ (*6,7*), $CrBr_3$ (*8, 9*), $CrCl_3$ (*10-12*), $Fe_3GeTe_2$ (*17,20*), and $RuCl_3$ (*21-23*). Out of these, only $Fe_3GeTe_2$ is a metallic ferromagnet and there is no known vdW-based 2D antiferromagnetic metal. Moreover, no evidence of high carrier mobilities has been reported in any of these exfoliated thin materials or even in their bulk vdW crystals. In general, high mobility is limited to very few vdW materials, such as graphite (*24*) and black phosphorus (*25*). A material with high electronic mobility and a corresponding high mean-free-path (MFP), might be critical for potential magnetic "twistronic" devices (*3*) where a large MFP could enable interesting phenomena in a Moiré-supercell induced flat band. In addition, conducting antiferromagnetic materials are the prime candidates for high-speed antiferromagnetic spintronic devices (*26*). Here we report the realization of a very high electronic mobility in a vdW layered antiferromagnet, $GdTe_3$, both in bulk and exfoliated thin flakes.

We chose to study $GdTe_3$, since rare-earth tritellurides ($RTe_3$, $R$ = La-Nd, Sm, and Gd-Tm) are structurally related to topological semimetal ZrSiS (*27,28*), while being known to exhibit an incommensurate charge density wave (CDW) (*29-31*), rich magnetic properties (*32*), and becoming superconducting under high-pressure ($R$ = Gd, Tb and Dy) (*33*). Combined, these properties hint that they could exhibit both high mobility and magnetism. Without considering the CDW, $R$Te$_3$ phases crystallize in an orthorhombic structure with the space group *Bmmb*. The structure is formed by double Te square-net sheets (perpendicular to the *c*-axis), separated by double corrugated $R$Te slabs, as illustrated in Fig. 1A. The $R$Te$_3$ crystal structure additionally exhibits a vdW gap between the two neighboring Te square-net sheets, which allows to exfoliate $R$Te$_3$ bulk crystals into 2D thin flakes.

High-quality plate-like GdTe$_3$ crystals, with lateral sizes up to ~8 mm × ~8 mm in the basal plane (inset in Fig. 1B), were grown by the self-flux technique (See Materials and Methods and Fig. S1 in Supplementary Materials). Despite the existence of a CDW along the *b*-axis below 379 K (*34*) (Fig. S2A), the GdTe$_3$ crystals show excellent metallicity. Under zero magnetic field, they reveal a large residual resistivity ratio (*RRR*, defined as $\rho_{xx}$(300 K)/$\rho_{xx}$(2 K)) reaching up to 825 (for a list of explored samples in this work see Table S1), therefore demonstrating excellent crystal quality with low defect concentrations. Figure 1B shows a representative scanning tunneling microscopy image of the cleaved GdTe$_3$ surface, indicating the CDW modulation (Fig. S3) and a very low defect concentration of one defect per 200 unit cells.

The antiferromagnetic order is confirmed by temperature-dependent DC magnetization of bulk GdTe$_3$ (Fig. 1C), which reveals three magnetic transitions: the Néel transition at $T_N$ = 12.0 K, and two further transitions at $T_1$ = 7.0 K and $T_2$ = 10.0 K; the $T_1$ transition has previously not been noted in the literature. Independently, these three transitions are confirmed by zero-field heat capacity, and temperature dependent resistivity measurements under an applied magnetic field (Fig. S4).

Quantum oscillation (QO) measurements demonstrate very long single-particle quantum lifetime in GdTe$_3$ crystals. Fig. 2A shows very detailed de Haas-van Alphen (dHvA) oscillations that appear in the AC magnetic susceptibility measurements. The fast Fourier transformation (FFT) of the dHvA oscillations (inset of Fig. 2A) reveals five oscillation frequencies: $F(\alpha) = 60$ T, $F(\beta_1) = 472$ T, $F(\beta_1) = 506$ T, $F(\gamma_1) = 813$ T and $F(\gamma_2) = 847$ T. By performing the Lifshitz-Kosevich (L-K) analysis (Fig. 2B) on the three strong QO components ($\beta_1$, $\beta_2$ and $\gamma_2$), the quantum lifetime is determined to be $13.5 - 17.5 \times 10^{-14}$ s, resulting in a mobility of 1400-1700 cm$^2$ V$^{-1}$s$^{-1}$ (for more details see Table S2).

Besides dHvA oscillations, Shubnikov–de Haas (SdH) oscillations, which appear in the field-dependent resistivity measurement (Figs. S6 and S7), provide an additional insight to the overall Fermi Surface (FS) geometry. Although different samples show slight difference in the relative peak intensities in the FFT spectra (see Fig. 2C and Fig. S6B), the FS pockets revealed by SdH oscillations are overall in good agreement with those determined from dHvA oscillations, with additional FS pockets ($\delta_1$ and $\delta_2$) appearing at higher frequencies of 3708 and 3948 T, respectively (Fig. S6B, also see Table S2 for comparison). Furthermore, the higher-frequency oscillations ($\beta$, $\gamma$ and $\delta$) are much weaker than the $\alpha$ oscillation and its harmonics. Possibly due to this reason, the weak $\gamma_1$ is not resolved in the SdH measurements.

The dominant $\alpha$ oscillation in SdH oscillations allows for the evaluation of the quantum lifetime of $\alpha$ FS pocket. Unlike the larger FS pockets (Fig. 2B), the amplitude of $\alpha$ oscillation (Fig. 2D) shows clear deviation from the L-K formula and appears to reach a plateau in the magnetically ordered regime ($T < T_N$). This suggests a significant interplay of the magnetic order to the conducting electrons in the small $\alpha$ pocket, but a negligible effect of the magnetic order in the larger FS pockets. L-K fits of the temperature dependence of the oscillation amplitude above $T_N$ yield a very light cyclotron effective mass of $m^*(\alpha) = 0.106 m_e$ ($m_e$ is the free electron mass) and a quantum lifetime up to $12.1 \times 10^{-14}$ s, resulting in a mobility of 2012 cm$^2$ V$^{-1}$ s$^{-1}$ (see Table

S2). The observed pockets in QO studies reasonably match with previously reported calculated Fermi surfaces of $R$Te$_3$ phases (*30*). A detailed discussion on the FS geometry is given in the supplementary text. We notice that the SdH oscillations in GdTe$_3$ have previously been reported in Ref. (*35*). However, in this study, only the $\alpha$ pocket was resolved, resulting in F($\alpha$) ≈ 56 T, and $m^*$ ($\alpha$) ≈ 0.1$m_e$. Such result on the $\alpha$ FS pocket is consistent with our measurement.

For high-speed device applications, it is crucial to determine the transport mobility and not just the mobility derived from the quantum lifetime. Hall measurements provide an important overview on the transport mobility and carrier concentration. Figures 3A,B show their temperature dependence, based on fits to the Hall resistivity, $\rho_{yx}$, assuming a two-band model. At 2 K, the electron and hole concentrations are evaluated to be $1.0 \times 10^{21}$ cm$^{-3}$ and $2.5 \times 10^{21}$ cm$^{-3}$, respectively, and the electron and hole transport mobilities are determined to be $\mu_t$ (e) = 28,100 cm$^2$ V$^{-1}$ s$^{-1}$ and $\mu_t$ (h) = 8,300 cm$^2$ V$^{-1}$ s$^{-1}$, respectively (also see Table 1). For accuracy, a two-band model fit is also performed on the low-field Hall conductivity, $\sigma_{xy}$ (see Materials and Methods in Supplementary Materials). It results $\mu_t$ (e) = 37,700 cm$^2$ V$^{-1}$ s$^{-1}$ and $\mu_t$ (h) = 13,500 cm$^2$ V$^{-1}$ s$^{-1}$, respectively (see Table 1). The results from these two methods reasonably agree with each other. Surprisingly, the electron transport mobility is found to be more than 14 times larger than the mobility ($\mu_q$) estimated from QO (Table S2). Such a large difference is uncommon, despite their different scattering processes (*36*). In the three-dimensional Dirac semimetal Cd$_3$As$_2$, for example, $\mu_t$ has been shown to be more than 12 times higher than $\mu_q$ (*36*). We surmise band topology might play a role in GdTe$_3$ as well. Among all measured samples, the highest achieved electron and hole mobilities are $\mu_t$ (e) = 61,200 − 113,000 cm$^2$ V$^{-1}$ s$^{-1}$ and $\mu_t$ (h) = 15,000 – 23,500 cm$^2$ V$^{-1}$ s$^{-1}$, respectively (Fig. S9). The lower value corresponds to the value obtained from the Hall conductivity fit, while the higher values is obtained from the Hall resistivity fit (See more details in Table 2).

While the transport mobility might seem low compared to some nonmagnetic topological semimetals, such as $Cd_3As_2$ (mobility of $9 \times 10^6$ $cm^2$ $V^{-1}$ $s^{-1}$ at 5 K (*36*)), it is among the highest reported for any magnetic compound. For reference, Table 2 presents a list of known layered magnetic materials (that cannot necessarily be exfoliated) with high mobility. Also listed is the structurally related nonmagnetic topological semimetal ZrSiS (*28*), as well as high-mobility delafossite $PdCoO_2$ (*37*), graphite (*24*) and black phosphorus (*25*). $GdTe_3$ stands out as the material with the highest mobility in the category of (quasi-)layered magnetic materials, and even compares to nonmagnetic black phosphorus. Among all the listed magnetic materials, $GdTe_3$ is the only vdW layered material and thus the only one that can be exfoliated.

The reason for such a high mobility in a magnetically ordered phase is closely related to the crystal structure of $GdTe_3$. Materials containing a square net as a structural motif frequently exhibit very steeply dispersed bands (*38*). In general, high mobility ($\mu$) is described by two factors in a conducting material: a low effective mass, $m^*$, and a long scattering time, $\tau$. While the long scattering time benefits from the high crystal quality, the low effective mass is related to the steep conduction bands resulting from the square net. To confirm the latter, we performed ARPES measurements (Fig. S13) and extracted the Fermi velocity ($V_F$) of the bands composing the pockets around $\bar{X}$ to be $1.1\sim1.2\times10^6$ m $s^{-1}$, which is a slightly conservative value compared to that of other $R$$Te_3$ compounds ($1.5\times10^6$ m $s^{-1}$) (*39*). Nevertheless, these numbers are of the same order as graphene (*40*). Besides the square-net structure, the CDW might also play an important role. The scattering time can be enhanced due to partial gap opening of the FS.

Finally, we provide evidence for the exfoliation capability of GdTe$_3$ and the persistence of high electronic mobility and magnetic order in a thin flake. Using a micromechanical exfoliation approach in inert atmosphere, we first created a 22-nm thin device of GdTe$_3$. Note that the thin flakes are degraded with heating or long exposure to air (Fig. S14). To ascertain that magnetic order is retained in such thin flakes, we performed transport measurements on the 22-nm flake. The temperature-dependent resistivity (Fig 3C) under an applied magnetic field shows magnetic phase transitions, visible in the slope change of the resistivity, similar to the bulk (Fig. S4B). Therefore, we conclude that the magnetic order still exists in such thin flakes. Based on the thin-flake transport measurements, the electron and hole mobility are determined to be 5700 cm$^2$ V$^{-1}$ s$^{-1}$ and 3300 cm$^2$ V$^{-1}$ s$^{-1}$, respectively (Fig. 3D). While these values are lower than in the bulk, the mobility is still slightly higher than that reported for black phosphorus flakes with similar thickness (*41*). The lower mobility in the GdTe$_3$ flake compared to the bulk might be related to slight sample degradation possibly due to short exposure to air, which results in a lower *RRR* of 67. Nonetheless, it is very likely that future effort that targets improvements of the thin-flake device fabrication, with, for example h-BN encapsulation (*42*), will enhance the mobility and realize the full potential of this material. We would also like to point out that the mobility of graphene in early studies (*43*) was much lower (3000 to 10,000 cm$^2$ V$^{-1}$ s$^{-1}$) than that of bulk graphite (*24*), but has been vastly improved to ~1,000,000 cm$^2$ V$^{-1}$ s$^{-1}$ (*44*) since. There are some additional observations that can be made from the transport experiment on the thin flake. For once, the resistivity measurement (Fig. 3C) suggests the persistence of the CDW, however, the transition temperature is slightly higher than in the bulk. Therefore, GdTe$_3$ may additionally provide a platform to study the thickness dependence of CDWs. Furthermore, the

SdH measurements on a 19-nm flake (Fig. S8) hint to a reduction of the FS pockets in size with sample thickness.

We extended the exfoliation procedure and were able to reach GdTe$_3$ flakes with a minimum thickness of 3.8 nm (Fig. 4A,B), which corresponds to three monolayers (where a monolayer is half a unit cell). The structural integrity (on flakes down to 7.5 nm) was confirmed by Raman spectroscopy measurements in inert atmosphere (Fig. S15). Therefore, we are confident that mono- or bilayer devices of GdTe$_3$ will be accessible with further optimized exfoliation conditions.

In summary, we showed that the layered antiferromagnetic compound GdTe$_3$ exhibits high electronic mobility. It stands out as the material that shows the highest mobility within all known layered magnetic materials, to the best of our knowledge. Among all known vdW materials, the dominant carrier mobility of GdTe$_3$ is only surpassed by graphite, but comparable to black phosphorus. We also demonstrated that GdTe$_3$ can be exfoliated to ultrathin flakes, reaching three monolayers (or 1.5 unit cells). In a GdTe$_3$ flake of ~20 nm, a relatively high carrier mobility is maintained, while the CDW and magnetic order persists. Overall, GdTe$_3$ can be considered to be the first high-mobility, magnetic van der Waals material. The combination of these properties provides a huge potential for novel 2D spintronic or twistronic devices. We believe that the establishment of a thin magnetic material with high mobility provides numerous exciting opportunities for future studies.

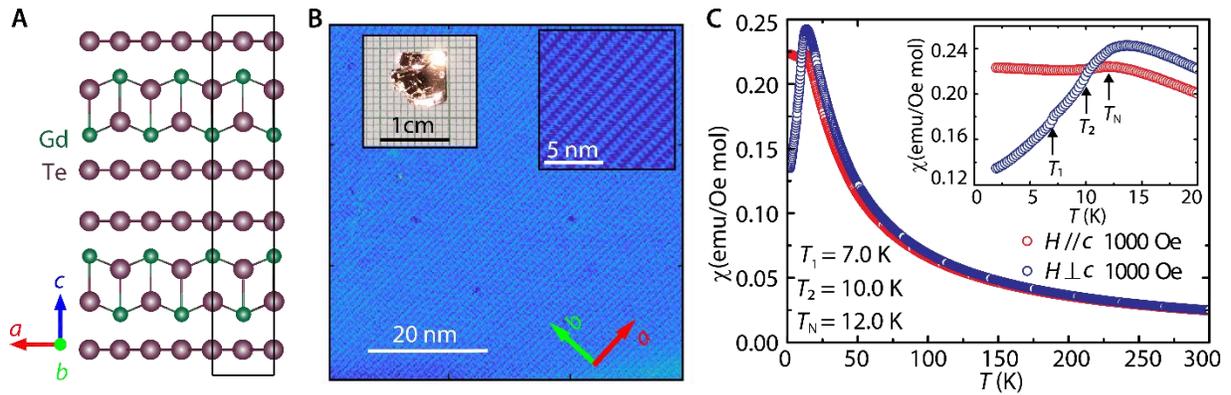

**Fig. 1. GdTe$_3$ crystal structure and antiferromagnetism.** (**A**) Illustration of the GdTe$_3$ crystal structure: a van der Waals gap is located between the two neighboring Te sheets. The rectangular box indicates the unit cell if no CDW is considered. (**B**) STM image of the GdTe$_3$ surface at 72 K with a tip bias of 0.2 V. The CDW vector is along $b$-axis. The left inset shows a typical GdTe$_3$ crystal. The right inset shows a zoom-in image with atomic resolution. (**C**) Temperature dependent magnetization of a bulk GdTe$_3$ crystal under zero-field-cooling conditions. $H//c$ and $H\perp c$ indicate the applied field perpendicular and parallel to the basal plane, respectively. The arrows indicate the three transitions.

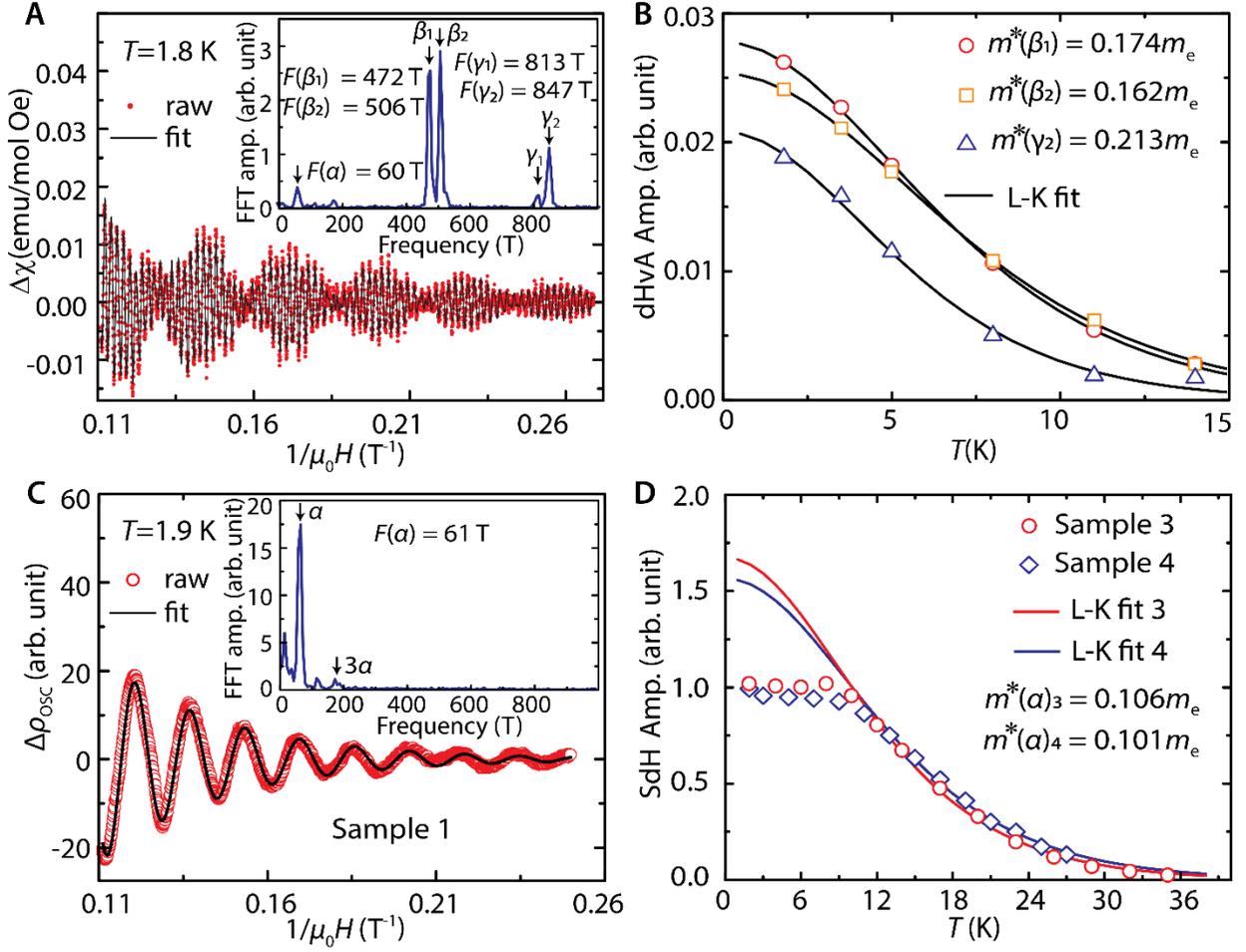

**Fig. 2. Quantum oscillations of bulk GdTe$_3$.** (**A**) dHvA oscillations at 1.8 K with L-K fit. The inset shows the FFT spectrum, with five indicated oscillation frequencies. (**B**) Temperature dependence of the amplitudes of $\beta_1$, $\beta_2$ and $\gamma_2$ oscillations from dHvA measurements. The solid lines are fits to the L-K formula. (**C**) SdH oscillations after subtracting the polynomial background from field-dependent resistivity measurements ($\rho_{xx}$) for Sample 3. The inset shows the FFT spectrum, with resolved $\alpha$ oscillation and its third harmonics. (**D**) Temperature dependence of the amplitude of the $\alpha$ oscillation from SdH measurements. The solid line is a fit to the L-K formula above $T_N$, resulting the effective mass of $m^*(\alpha)_3$ and $m^*(\alpha)_4$ for Sample 3 and 4, respectively.

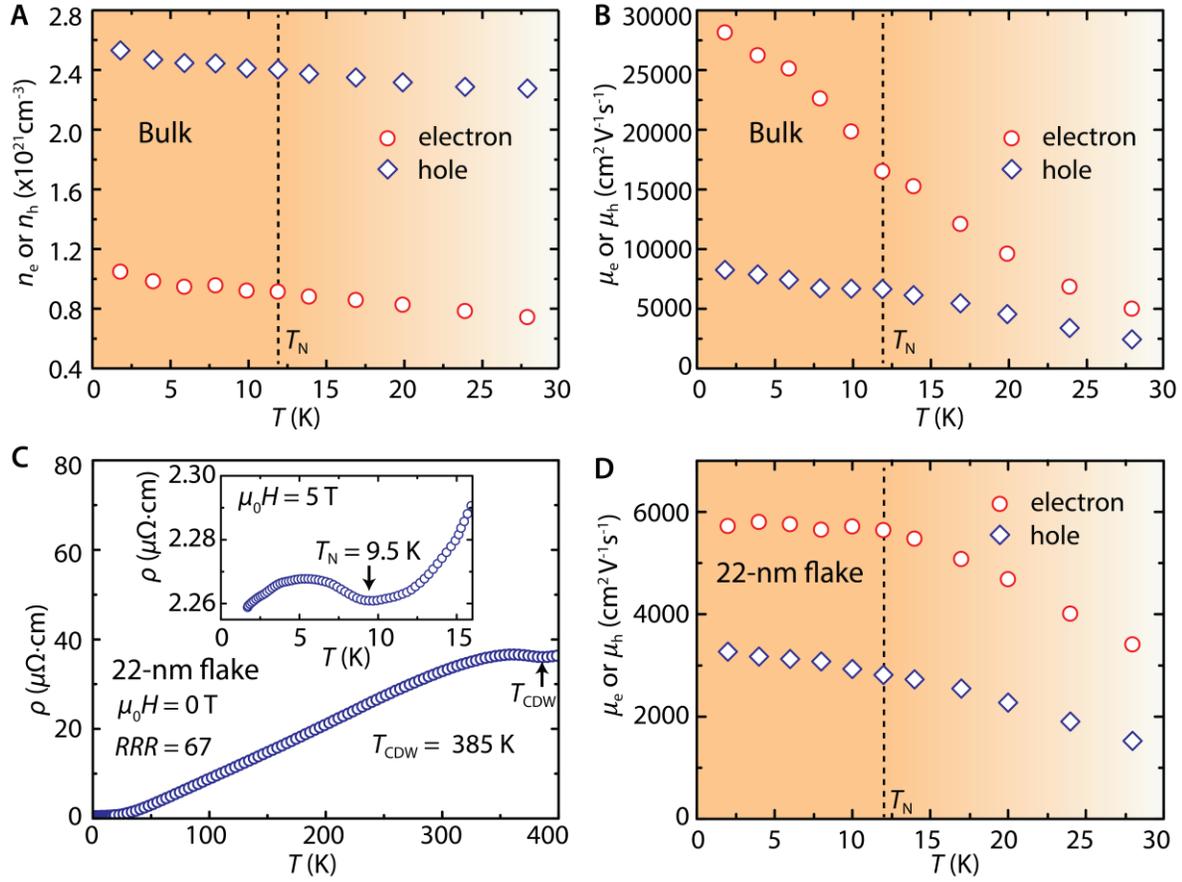

**Fig. 3. Carrier concentrations and transport mobilities of bulk GdTe$_3$ and a 22-nm flake.**

(**A,B**) Temperature-dependent carrier concentrations and mobilities from Hall measurements of bulk GdTe$_3$. The dashed lines indicate $T_N$. (**C**) Temperature dependent resistivity on a 22-nm thin flake, showing both the existence of the CDW and the AFM transition. The inset shows the low-temperature resistivity under an applied field of 5 T, revealing the magnetic transition. (**D**) Temperature dependent electron and hole mobilities of the 22-nm thin flake.

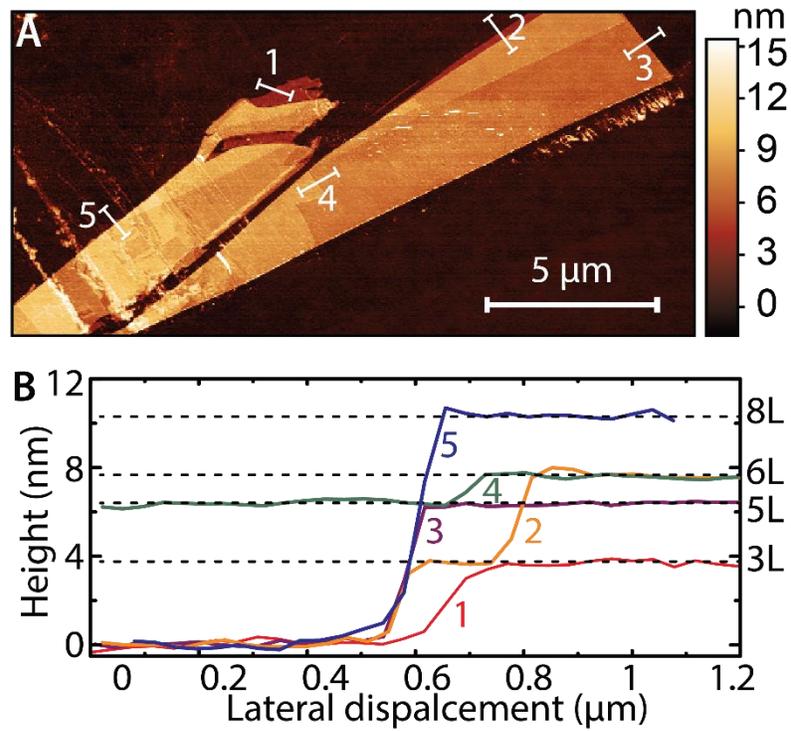

**Fig. 4. Exfoliation of GdTe$_3$ ultrathin flakes. (A,B),** An AFM image of exfoliated GdTe$_3$ ultrathin flakes and its cross-sectional height profiles. Note that the height profiles are translated into the number of GdTe$_3$ layers on the right vertical axis in **B**. One layer corresponds to half a unit cell (shown in Fig. 1A).

**Table 1**. **Carriers concentrations and mobilities from Hall measurements.** The results outside and inside the parentheses are from fits to the Hall resistivity ($\rho_{xy}$) and Hall conductivity ($\sigma_{xy}$), respectively.

| Sample Number | $n_e$ ($\times 10^{21}$ cm$^{-3}$) | $n_h$ ($\times 10^{21}$ cm$^{-3}$) | $\mu_t(e)$ (cm$^2$/Vs) | $\mu_t(h)$ (cm$^2$/Vs) | Sample Geometry |
|---|---|---|---|---|---|
| 1 | 1.05 | 2.53 | 28,100 | 8300 | Bulk |
|   | (0.61) | (2.03) | (37,700) | (13,500) | |
| 3 | 0.96 | 2.41 | 17,700 | 6000 | |
|   | (0.62) | (2.02) | (23,300) | (8400) | |
| 4 | 1.07 | 2.70 | 14,000 | 5100 | |
|   | (1.15) | (2.75) | (12,000) | (5,400) | |
| 5 | 1.59 | 2.74 | 113,000 | 15,000 | |
|   | (2.28) | (3.43) | (61,200) | (23,500) | |
| 6 | 1.01 | 2.15 | 5700 | 3300 | Thin flake |
|   | (1.12) | (2.05) | (5400) | (3300) | |

**Table 2.** A compilation of magnetic bulk materials, in addition to ZrSiS, PdCoO$_2$, graphite and black phosphorus, for which high mobilities are reported, in comparison to GdTe$_3$. For the transport mobility ($\mu_t$) estimated from Hall measurement, the values outside and inside the parentheses represent the electron and hole carriers, respectively. For the quantum lifetime derived mobility ($\mu_q$) and effective mass ($m^*$) estimated from SdH and dHvA oscillations, a range with lower and upper bounds is provided. For the transport mobility estimated from a combination of the QO and residual resistivity measurement, we denote it as "hybrid". The transport mobility estimated from magnetoresistance (MR) is listed when it is considered to be more accurate than the Hall mobility. The mobilities of PdCrO$_2$ and PdCoO$_2$ were deduced by the hybrid method because no quantum lifetime or Hall carrier mobility is reported in the literature. The mobility of EuMnBi$_2$ from the hybrid method is also listed for comparison with the Hall carrier mobility.

| Material | $\mu_q$ or $\mu_t$ (cm$^2$ V$^{-1}$ s$^{-1}$) | $m^*/m_e$ | Method | Reference |
|---|---|---|---|---|
| SrMnBi$_2$ | 250 | 0.29 | SdH | 45 |
| CaMnBi$_2$ | 488 | 0.53 | SdH | 46 |
| Sr$_{1-y}$Mn$_{1-z}$Sb$_2$ | 570 | 0.04-0.05 | SdH | 47 |
| YbMnBi$_2$ | 689 | 0.27 | SdH | 48 |
| YbMnSb$_2$ | 1584 | 0.134 | SdH | 49 |
| | 1072 | 0.108 | dHvA | 49 |
| | 6538 (1310) | NA | Hall | 49 |
| EuMnBi$_2$ | 1.6 (520) | NA | Hall | 50 |
| | (14,000) | NA | hybrid | 51 |
| BaMnSb$_2$ | 1280 | 0.052-0.058 | SdH | 52 |
| | 1300 | NA | Hall | 52 |
| GdPtBi | not reported | 0.23 | SdH | 53 |
| | (1500–2000) | NA | Hall | 53 |
| PdCrO$_2$ | 8700 | 0.33-1.55 | hybrid | 54 |
| BaFe$_2$As$_2$* | 1130 | NA | MR | 55 |
| GdTe$_3$ | 1165-2012 | 0.101-0.213 | SdH and dHvA | This work |
| | 134,000 (14,000) | NA | Hall | |
| ZrSiS | 1300-6200 | 0.1-0.14 | SdH | 28 |
| | 4219-10,000 | 0.025-0.052 | dHvA | 56 |
| | 20,000 (2800) | NA | Hall | 28 |
| PdCoO$_2$ | 51,000 | 1.45-1.53 | hybrid | 37 |
| graphite† | 1,263,000 | NA | MR | 24 |
| black phosphorus‡ | (65,000) | NA | Hall | 25 |

*The average mobility value from Hall data is 376 cm$^2$ V$^{-1}$ s$^{-1}$, but it was considered to be inaccurate. Therefore, the MR was used to evaluate the average mobility.
† The average mobility is adopted.
‡ The hole mobility in p-type black phosphorus is adopted as it is higher than the electron mobility n-type one.

**Acknowledgements**

This work was supported by NSF through the Princeton Center for Complex Materials, a Materials Research Science and Engineering Center DMR-1420541. LMS and SL were additionally supported by a MURI grant on Topological Insulators from the Army Research Office, grant number ARO W911NF-12-1-0461. The device fabrication was performed in part at the PRISM clean room at Princeton University. JL, TG and NPO acknowledge the support from the US Department of Energy (contract DE SC0017863). GF and AY acknowledge the support of ExxonMobil through Andlinger Center for Energy and the Environment. MG and KSB acknowledge support from the National Science Foundation under grant DMR-1410428. AT was supported by the DFG; proposal no. SCH 1730/1-1. This research used resources of the Advanced Photon Source, a U.S. Department of Energy (DOE) Office of Science User Facility operated for the DOE Office of Science by Argonne National Laboratory under Contract No. DE-AC02-06CH11357; additional support by National Science Foundation under Grant no. DMR-0703406. We thank B.A. Bernevig for helpful discussions.


**Author contributions**

SL and LMS initiated the project by identifying $GdTe_3$ as a potential high mobility magnetic 2D material. SL grew the crystals with help from SK and LMS. SL performed bulk DC and AC magnetization measurements, dvHA oscillation analysis and the heat capacity measurements. SL and JL performed the transport measurements and analyzed the data with input from TG, NPO and LMS. GF and AY investigated the crystals with STM. ARPES measurements were performed by AT, SL, JM and FR and were interpreted by AT, CRA, SL and LMS. The initial exfoliation tests with Raman spectroscopy and AFM inside an Argon-filled glove box were

performed by SL, MG and KSB. Thin-flake device fabrication and characterization were performed by YJ, SL, JL and SW. Air sensitivity tests of GdTe$_3$ thin flakes were performed by YJ and SW. The manuscript was written by SL and LMS with input from all authors.

**Competing interests** The authors declare no competing interests.

**Data availability** All data needed to evaluate the conclusions of the paper are present in the paper and/or the Supplementary Materials. Additional data related to this paper may be requested from the authors.

**Supplemental Materials**

Materials and Methods

Supplementary text

Fig. S1-S15

Table S1-S2

# Supplementary Materials for

# High mobility in a van der Waals layered antiferromagnetic metal


Shiming Lei[1], Jingjing Lin[2], Yanyu Jia[2], Mason Gray[3], Andreas Topp[4], Gelareh Farahi[2], Sebastian Klemenz[1], Tong Gao[2], Fanny Rodolakis[5], Jessica L. McChesney[5], Christian R. Ast[4], Ali Yazdani[2], Kenneth S. Burch[3], Sanfeng Wu[2], N. Phuan Ong[2] and Leslie M. Schoop[1]*

[1]*Department of Chemistry, Princeton University, Princeton, New Jersey 08544, USA*

[2]*Department of Physics, Princeton University, Princeton, New Jersey 08544, USA*

[3]*Department of Physics, Boston College, Boston, MA 02467, USA*

[4]*Max-Planck-Institut für Festkörperforschung, Heisenbergstraße 1, D-70569 Stuttgart, Germany*

[5]*Argonne National Laboratory, 9700 South Cass Avenue, Argonne, Illinois 60439, USA*

Email: lschoop@princeton.edu


**This PDF file includes:**

Materials and Methods

Supplementary Text

Figs. S1 to S15

Table S1 to S2

References



I. Materials and Methods

1.1 GdTe$_3$ crystal growth and x-ray diffraction

High-quality GdTe$_3$ single crystals were grown in an excess of tellurium (Te) via a self-flux technique. Te (metal basis >99.999%, Sigma-Aldrich) was first purified to remove oxygen contaminations and then mixed with gadolinium (Gd) (>99.9%, Sigma-Aldrich) in a ratio of 97:3. The mixture was sealed in an evacuated quartz ampoule and heated to 900 °C over a period of 12 hours and then slowly cooled down to 550 °C at a rate of 2 °C/hour. The crystals were obtained by a decanting procedure in a centrifuge. The as-grown crystals were characterized by x-ray diffraction using a Bruker D8 Advance Eco diffractometer in reflection geometry with Cu K$\alpha$ radiation and a STOE STADI P diffractometer in transmission geometry with Mo K$\alpha$ radiation. The room-temperature lattice parameters evaluated this way are 4.316 Å, 4.325 Å, and 25.6 Å along *a*-, *b*- and *c*-axis respectively, without considering the CDW modulations. For the exfoliated samples, we call a flake "*ultrathin*" when it is less than 3 unit cell thick, while it is "thin" when it is above 3 unit cell thick and below ~100 nm.

1.2 STM imaging of GdTe$_3$

Samples were cleaved and measured at 72 K in ultrahigh vacuum with a variable-temperature STM. A mechanically sharpened platinum-iridium tip was used for topography and spectroscopy measurements. The STM tip was treated on a Cu (111) surface before the experiment to ensure its metallic character. STM spectra were taken using a lock-in amplifier with setpoint current and bias of 60 pA and 800 mV respectively, and a modulation bias of 8 mV.

1.3 Magnetization and de Haas-van Alphen (dHvA) measurements

Temperature dependent DC magnetization measurements were performed on a Quantum Design PPMS DynaCool system via the Vibrating Sample Magnetometer (VSM) option. The AC magnetic susceptibility measurements were performed via the ACMS option. The dHvA oscillation was obtained by subtracting a polynomial background from the AC susceptibility data. A Fast Fourier Transformation (FFT) was performed to reveal the frequency of each quantum oscillation component. In order to quantify the amplitude of the three dominant QO components of $\beta_1$, $\beta_2$ and $\gamma_2$, the dHvA oscillation was fitted with a superposition of three exponentially decaying sinusoidal functions (*30*). The amplitude of each fitted QO component at the envelope peak location of 7.8 T was adopted for the evaluation of their respective cyclotron effective mass. For the dHvA oscillation, the temperature dependent QO damping is described by the Lifshitz-Kosevich (LK) formula:

$$\Delta M \propto -B^{1/2} \frac{\lambda T}{\sinh(\lambda T)} e^{-\lambda T_D} \sin\left[2\pi\left(\frac{F}{B} - \frac{1}{2} + \beta + \delta\right)\right], \tag{S1}$$

where $\lambda = (2\pi^2 k_B m^*)/(\hbar e B)$, $k_B$ is Boltzmann constant, $\hbar$ is reduced Planck's constant, $T_D$ is Dingle temperature, $F$ is QO frequency and $2\pi\beta$ is the Berry phase. In equation (S1), $\delta$ is a phase shift, which is 0 and $\pm\frac{1}{8}$ for the 2D and 3D systems, respectively. Because of the dominant contribution of the sinusoidal function in equation (S1) to the oscillatory component of the AC magnetic susceptibility ($\Delta\chi$), $\Delta\chi$ can be derived as:

$$\Delta\chi \propto -B^{-3/2} F \frac{\lambda T}{\sinh(\lambda T)} e^{-\lambda T_D} \cos\left[2\pi\left(\frac{F}{B} + \beta + \delta\right)\right]. \tag{S2}$$

In equation (S2), a thermal factor can be defined as:

$$R_T = \frac{\lambda T}{\sinh(\lambda T)}, \tag{S3}$$

to describe the temperature dependent damping of the dHvA oscillations. At a constant field, this thermal factor only depends on the cyclotron effective mass and temperature. The effective

masses of QO components of $\beta_1$, $\beta_2$ and $\gamma_2$ were thus determined. The following step involves a refitting of the dHvA oscillation at 1.8 K to a superposition of three QO components, as described by equation (S2). The Dingle temperature ($T_D$) was then determined. The quantum lifetime ($\tau_q$) and mobility ($\mu_q$) were derived by $\tau_q = \hbar/2\pi k_B T_D$ and $\mu_q = e\tau_q/m^*$, respectively.

1.4 Electronic characterizations and SdH oscillations of Bulk GdTe$_3$

In-plane magnetoresistance (MR) measurements were carried out in a standard four-terminal geometry in a Quantum Design PPMS DynaCool system. A constant AC current with amplitude of 10 mA was applied. For the evaluation of the SdH oscillations, a polynomial background was subtracted from the symmetrized MR data. The cyclotron effective mass and Dingle temperature of the $\alpha$ FS pocket were determined in a similar manner as that in dHvA oscillation, except in this case only the $\alpha$ FS pocket is evaluated. The SdH oscillation is described by the L-K formula:

$$\Delta\rho \propto \frac{\lambda T}{\sinh(\lambda T)} e^{-\lambda T_D} \cos\left[2\pi\left(\frac{F}{B} - \frac{1}{2} + \beta + \delta\right)\right]. \tag{S4}$$

Note that the results of SdH oscillations from two samples (#3 and #4) are shown in Fig. 2D. Sample 3 has a *RRR* of 315, while Sample 4 has a *RRR* of 188. Despite the difference in *RRR*, the effective mass of the $\alpha$ FS pocket determined on these two samples is very close, as shown in Fig. 2D. The angle dependent SdH oscillations were measured with the aid of PPMS Horizontal Rotator. For the evaluation of the transport mobility and carrier concentrations, Hall measurements ($\rho_{xx}$ and $\rho_{yx}$) were performed in a standard Hall bar geometry. Assuming a two-band model, the Hall resistivity is expressed as (*57*):

$$\rho_{yx}(B) = \frac{B}{e} \frac{(n_h\mu_h^2 - n_e\mu_e^2) + (n_h - n_e)(\mu_e\mu_h B)^2}{(n_e\mu_e + n_h\mu_h)^2 + (n_h - n_e)^2(\mu_e\mu_h B)^2}, \tag{S5}$$

where $B = \mu_0 H$ ($\mu_0$ is the free-space permeability), e is the elementary charge, $n_e$ and $n_h$ are electron and hole carrier concentration, respectively, and $\mu_e$ and $\mu_h$ are the electron and hole carrier mobility, respectively. According to the two-band model, the zero-field resistivity, $\rho_{xx}(0)$, is related to the carrier concentration and mobility through the following equation:

$$\rho_{xx}(0) = \frac{1}{e} \frac{1}{n_e \mu_e + n_h \mu_h}. \tag{S6}$$

The carrier concentration and mobility were determined by fits of the Hall resistivity to equation (S5) on the condition of equation (S6).

Based on equation (S5), one can see that the field dependent Hall resistivity becomes linear at high fields; the slope is:

$$\frac{d\rho_{yx}}{dB}\bigg|_{B \text{ is large}} = \frac{1}{(n_h - n_e)e}, \tag{S7}$$

which solely depends on the difference in carrier concentration. Such behavior is clearly observed in our transport measurements on GdTe$_3$. For completion, fits to the two-band model are also performed on the Hall conductivity ($\sigma_{xy}$) data which is converted from $\rho_{xx}$ and $\rho_{yx}$. $\sigma_{xy}$ is related to the carrier concentration and mobility through the expression:

$$\sigma_{xy}(B) = eB \left( \frac{n_h \mu_h^2}{1 + \mu_h^2 B^2} - \frac{n_e \mu_e^2}{1 + \mu_e^2 B^2} \right). \tag{S8}$$

While the high-field Hall resistivity provides information on the carrier concentration difference, the low-field Hall conductivity is sensitive to the specific carrier type that has a higher mobility, even if it is the minority carrier. The Hall conductivity fits are performed with the carrier concentration difference and the zero-field resistivity. The mobilities evaluated from the Hall resistivity and Hall conductivity fits respectively, are considered as the two boundaries. The real carrier mobilities are considered to lie in between.

1.5 GdTe$_3$ thin flake exfoliation and device fabrication.

GdTe$_3$ crystals were micromechanically exfoliated using scotch tape and deposited directly onto 285 nm SiO$_2$/Si substrates. We performed the exfoliation tests both inside and outside the inert gas glovebox. The thin flakes were identified under an optical microscope based on the optical contrast on the substrate. Their thickness and structural integrity were directly measured in inert atmosphere through an NMI ezAFM40, and Witex alpha300 Raman microscope. For Raman, a 100x objective was employed with a 532nm laser. To avoid damage and overheating the laser power was kept below 50 µW. For thin-flake device fabrications, the pre-patterned Hall bar electrode (25 nm Au with 5 nm Ti as the sticking layer) was deposited on a 285 nm SiO2/Si substrate by standard E-beam lithography and evaporation procedure. In order to achieve a smooth top surface of the electrode, we used double-layer poly(methyl methacrylate) (PMMA) as the resist (PMMA 495 A2 as the bottom layer and 950 A2 as the top layer). Afterwards, GdTe$_3$ thin flakes were exfoliated onto a SiO2/Si substrate in an Argon-filled glovebox and transferred onto the pre-patterned electrode by standard dry transfer techniques for 2D materials, using a Poly (Bisphenol A Carbonate) (PC) / Polydimethylsiloxane (PDMS) stamp. To protect the flake from air exposure, the PC film was left on top of the device. The device was then transported to a PPMS for further electrical characterizations. In the whole process, the air-exposure was limited to less than 30 minutes. After the measurement, the PC film was dissolved in chloroform, and the thickness of thin flake was measured by AFM.

1.6 ARPES measurement on bulk GdTe$_3$

Soft x-ray ARPES measurements were performed at 9 K at the IEX beam line (29ID, Advanced Photon Source, Argonne National Laboratory) using a hemispherical Scienta R4000 electron analyzer with a pass energy of 200 eV (energy and angular resolution are 220 meV and 0.1°, respectively). The ARPES spectra were recorded with right circular polarized light at a

photon energy of 500 eV. GdTe$_3$ was cleaved cold, 30 K, and then cooled down to 9 K for the ARPES measurements.

II. Supplementary Text

2.1 Overview of the samples that were characterized by electrical transport in this work

**Table S1. An overview of the GdTe$_3$ samples (bulk and thin flake geometries), on which we have performed transport measurements in this work.** Note: The residual resistivity ratio (*RRR*) is defined in the main text as $\rho_{xx}$ (300 K)/$\rho_{xx}$ (2 K) from the in-plane resistivity ($\rho_{xx}$) measurements under zero magnetic field; the magnetoresistance (MR) is defined as the ratio of the change of $\rho_{xx}$ under magnetic field, $(\rho_{xx}(H) - \rho_{xx}(0))/\rho_{xx}(0) \times 100\%$, where $\rho_{xx}(0)$ is the resistivity at zero-field. The thickness of the bulk samples (Sample 1-5) ranges from 10 μm to 30 μm and the thin flake sample 6 and 7 are 22 nm and 19 nm thick, respectively. Although the *RRR* of thin flake sample 7 is not measured due to imperfect contacts during the cooling process, SdH oscillations can be clearly identified at low temperature (Fig. S8).

| Sample Number | *RRR* | MR (9 T) | Sample Geometry |
|---|---|---|---|
| 1 | 358 | 1400% | Bulk |
| 2 | 511 | 2800% | Bulk |
| 3 | 315 | 1300% | Bulk |
| 4 | 188 | 970% | Bulk |
| 5 | 825 | 2900% | Bulk |
| 6 | 67 | 650% | Thin flake (22 nm) |
| 7 | NA | NA | Thin flake (19 nm) |

2.2 Crystal quality evaluated by x-ray diffraction, SEM/EDX and zero-field resistivity

The GdTe$_3$ crystals were characterized by x-ray diffraction (XRD) and SEM/EDX to confirm the structure and composition (Fig. S1). Right before these characterizations, the crystal surface was cleaned by peeling off the exterior layers with scotch-tape. No impurity phases can be detected in the XRD or SEM/EDX characterizations.

The high quality of the GdTe$_3$ crystals is also supported by a high *RRR* in the in-plane resistivity measurements. The temperature-dependent zero-field resistivities on two representative GdTe$_3$ crystals (Sample 1 and 2) are shown in Fig. S2. The resistivity of Sample 1 was investigated up to 400 K to determine the CDW transition temperature of $T_{\mathrm{CDW}} = 379$ K. The *RRR*s of Sample 1 and 2 are determined to be 358 and 511, respectively. A fit to the quadratic temperature-dependency relation, $\rho(T)= \rho(T)+AT^2$, below $T_{\mathrm{N}} = 12$ K reveals a prefactor $A$ of 0.55 n$\Omega$ cm/K$^2$ and 0.58 n$\Omega$ cm/K$^2$, respectively. These values are only slightly higher than the typical values for transition metals (*58*), but much lower than that of Bi (7.7 n$\Omega$ cm/K$^2$) (*59*), SrMnBi$_2$ (19 n$\Omega$ cm/K$^2$) (*45*), and YbMnBi$_2$ (5.7 n$\Omega$ cm/K$^2$) (*48*). Since $A$ is inversely proportional to the Fermi temperature, the low prefactor $A$ indicates that light carriers are responsible for the metallic conduction (*45*).

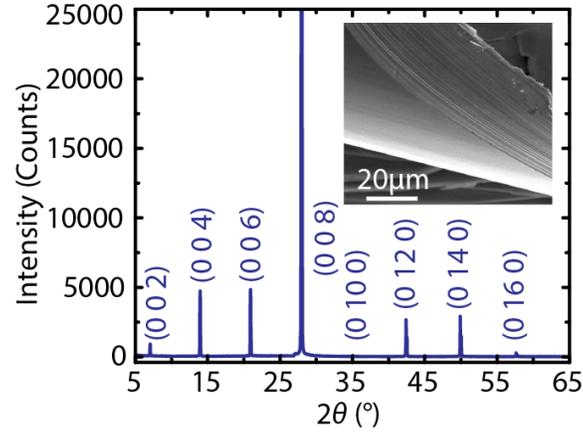

**Fig. S1. X-ray diffraction pattern and scanning electron microscopy (SEM) image of a GdTe3 crystal.** X-ray diffraction was measured on a plate-like single crystal so that only the (00*l*) diffractions are visible. The inset shows a side-view SEM image, highlighting the layering morphology of GdTe$_3$ crystals.

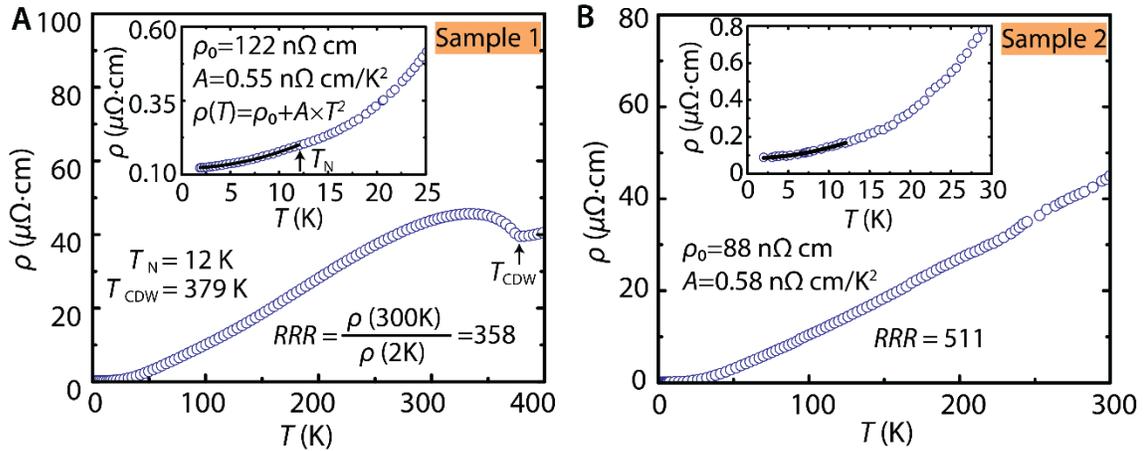

**Fig. S2. Temperature dependent in-plane resistivity of Sample 1 and 2.** (**A**) Sample 1 has *RRR* of 358 and the CDW transition is visible at $T_{CDW}$ = 379 K. The inset shows the low temperature resistivity. The black line is a fit to the quadratic temperature-dependent resistivity relation. (**B**) Sample 2 with a *RRR* of 511. The inset shows the low temperature resistivity with a quadratic temperature dependent fit (black line).

2.3 Scanning tunneling microscopy (STM) topography and spectroscopy

A Fourier analysis of the measured STM topographic image is able to reveal the CDW wave vector (*60,61*). Figures S3A, B show the Fourier transformed image with a cross-sectional cut to evaluate the CDW vector. The estimated CDW vector is $q_{CDW} \approx 2/7 \times 2\pi/b = 4.15$ nm$^{-1}$. The STM spectrum (dI/dV curve) shown in Fig. S3C is an averaged result over a small area, using a lock-in amplifier with setpoint current and bias of 60 pA and 800 mV respectively, with a modulation bias of 8 mV. The shape of the dI/dV curve is remarkably similar to those from previous measurements on TbTe$_3$ (*60*) and CeTe$_3$ (*61*). Since dI/dV is proportional to the local density of states (DOS), the reduced intensity in the dI/dV curve reflects the partial gap opening of the Fermi surface (FS). Based on the information from the dI/dV curve, the estimated energy width is $2\Delta_{CDW} \approx 420$ mV, and it gives the CDW gap of ~210 mV. This is similar to that observed TbTe$_3$ (*60*), and CeTe$_3$ (*61*). The finite conductance at zero bias results from the nonzero DOS inside the gap.

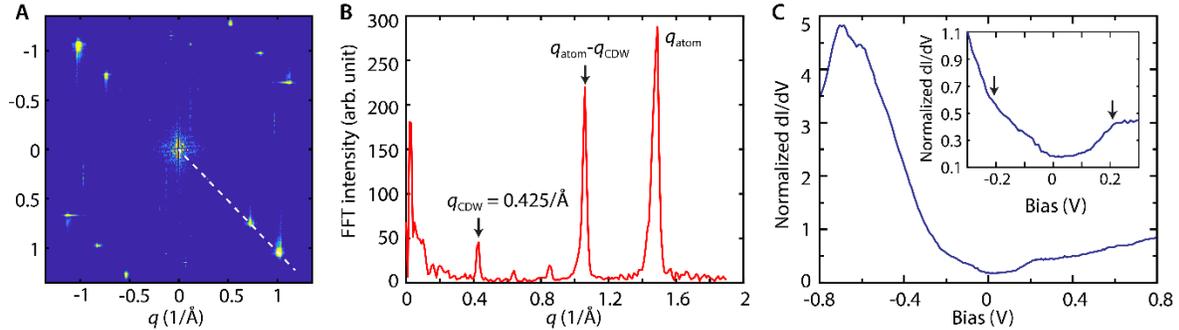

**Fig. S3. CDW revealed by STM on a GdTe$_3$ crystal.** (**A**) 2D FFT analysis of an STM topography image on a GdTe$_3$ crystal. (**B**) A cross-sectional line-cut (the path is indicated as a dashed line in **A**) of the FFT image. The CDW vector is estimated to be $q_{CDW} = 4.25$ nm$^{-1}$. (**C**) dI/dV curve. Inset shows the zoom-in spectrum. The arrows mark the edge location for the estimation of the CDW gap $\Delta_{CDW}$.

2.4 Magnetic transitions revealed by specific heat and resistivity

The three magnetic transitions observed in the magnetic susceptibility measurement (Fig. 1C in the main text) can be also seen in the zero-field specific heat measurement (Fig. S4A) and the resistivity measurement under an applied field (Fig. S4B). Note that the $T_1$ transition might be accompanied by a structural transition based on the peak shape in the specific heat and sharp step in the resistivity curve. The exact magnetic orderings of these observed magnetic ordered phases, however, are so far not clear.

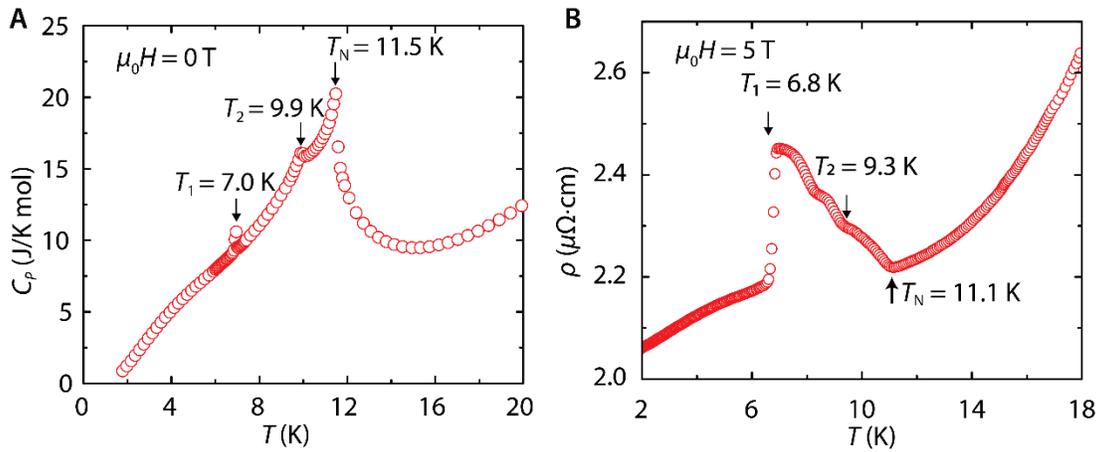

**Fig. S4. Heat capacity (A) and resistivity measurements (B) on GdTe$_3$ bulk crystal.** Note that the heat capacity is measured under zero magnetic field while the resistivity is measured under 5 T field, therefore the latter measured transition temperatures are systematically lower than the former ones. The three transitions at $T_1$, $T_2$ and $T_N$ are corresponding to the three transitions determined from the DC magnetization measurement (Fig. 1C in the main text).

2.5 Effective magnetic moment

A Curie-Weiss fit to the DC magnetic susceptibility on a GdTe$_3$ crystal is shown in Fig. S5. The measured ordered magnetic moment of 7.91 $\mu_B$/Gd agrees very well with the free ion value for Gd$^{3+}$ (7.94 $\mu_B$/Gd), thus suggesting that the magnetic moments are localized below the Fermi level. This suggests that the $R$Te slabs are magnetically localized and therefore have an insulating character, while the Te square-net sheets contribute to the metallic nature of GdTe$_3$.

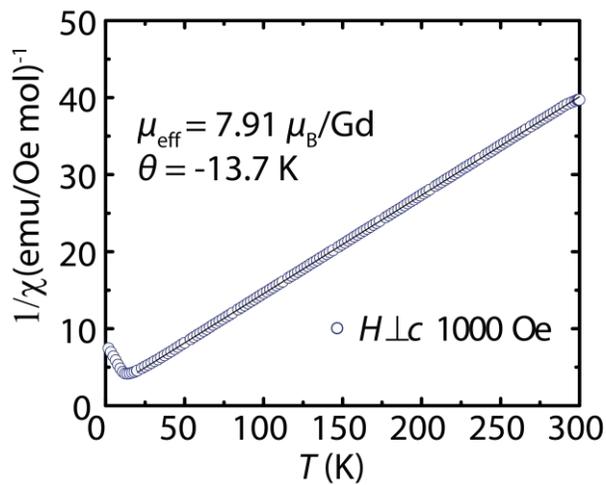

**Fig. S5. Curie-Weiss fit to the inverse magnetic susceptibility in the paramagnetic regime.** The Weiss temperature $\theta$, and the effective moment $\mu_{eff}$ are determined to be -13.7 K and 7.91 $\mu_B$/Gd, respectively. The effective moment $\mu_{eff}$ is in very close agreement to the theoretical value of 7.94 $\mu_B$/Gd for a localized Gd$^{3+}$ state with $S=7/2$.

2.6 Shubnikov–de Haas (SdH) oscillations

The field dependent in-plane resistivity was used to analyze the SdH oscillations. As mentioned in the main text, the SdH oscillations on GdTe$_3$ crystals generally show much weaker high-frequency oscillations compared to the dHvA measurements. Figures S6A, B shows the SdH oscillations on Sample 1 (after polynomial background subtraction) in the range of 6 T to 9 T and its FFT spectrum. Here the $\alpha$, $\beta$, $\gamma$ and $\delta$ pockets are clearly resolved in the FFT spectrum (Fig. S6B). The $\gamma_1$ frequency that was resolved in the dHvA measurement, was not resolved here, most likely due to its weaker intensity compared to $\alpha$. Note that in all samples where SdH oscillations were measured, the third harmonic 3$\alpha$ can be observed. In some samples with slightly lower *RRR* (such as Samples 3 and 4), the higher-frequency oscillations ($\beta$, $\gamma$ and $\delta$) cannot be resolved, and the third harmonic 3$\alpha$ oscillation also has a lower intensity. The dominant $\alpha$ oscillation in these samples (Samples 3 and 4) thus allows for an accurate evaluation of the cyclotron effective mass and quantum lifetime for the $\alpha$ pocket. A representative MR curve (Sample 3) is shown in Fig. S7A; it appears very similar to some nonmagnetic semimetals (*36*). It gives rise to a MR of 1300 % at 1.9 K and 9 T. The same in-plane resistivity measurements were performed under various sample tilt angles at 1.9 K. The resulting tilt angle dependent SdH oscillations are shown in Fig. S7B. The angle dependence of the $\alpha$ oscillation gives information about its FS dimensionality. With increasing $\theta$, the amplitude of the $\alpha$ oscillation decreases, while the oscillation frequency increases (inset of Fig. S7B). The angular dependent QO frequency nicely follows the factor of 1/cos$\theta$ up to 60°, suggesting a rather 2D FS morphology, in agreement with the layered crystal structure.

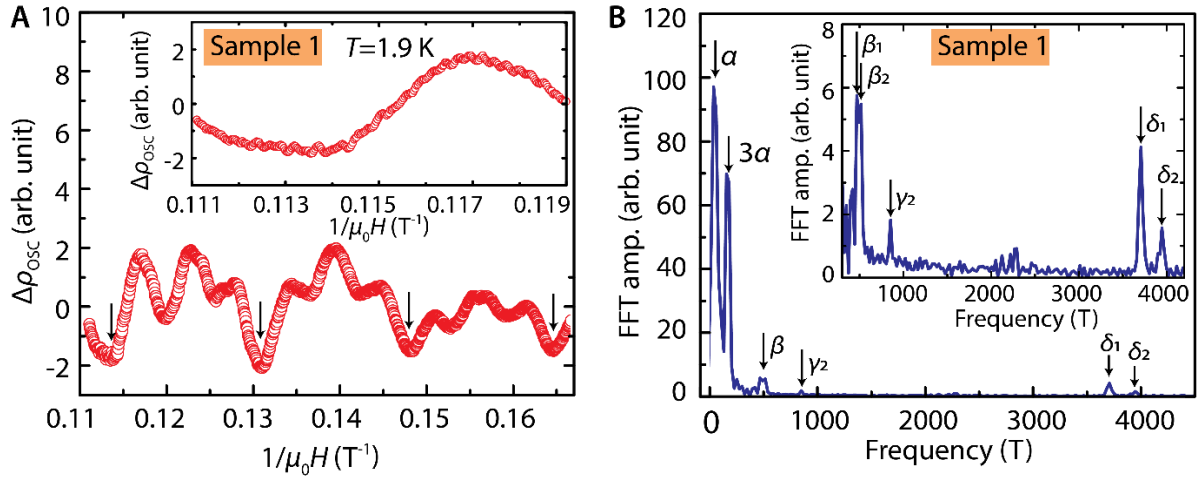

**Fig. S6. SdH oscillations ($H//c$) at 1.9 K that reveals detailed FS pockets.** (**A**) SdH oscillations in the range of 6 T to 9 T measured on Sample 1, the inset shows a zoom-in view of the quantum oscillations, revealing the high-frequency oscillations. The arrows indicate the valley of the $\alpha$ oscillation. The third harmonic $3\alpha$ oscillation is clearly revealed between two neighboring valleys. (**B**) FFT spectrum of the SdH oscillations shown in **A**. Besides the $\alpha$ pocket, it reveals the additional existence of the larger FS pockets $\beta_1$, $\beta_2$, $\gamma_2$ and $\delta_1/\delta_2$ (the $\delta$ pockets were not resolved in the dHvA measurements). The inset shows a zoom-in view of the FFT spectrum.

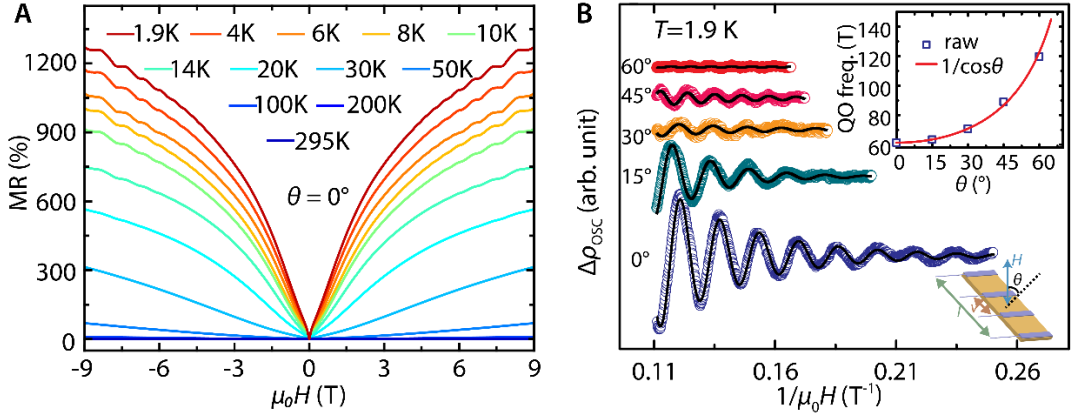

**Fig. S7. Field-dependent MR and angle-dependent SdH oscillations.** (**A**) MR measured at various temperatures for the field perpendicular ($\theta = 0°$) to the GdTe$_3$ layering plane. MR = ($\rho_{xx}(H) - \rho_{xx}(0))/\rho_{xx}(0) \times 100\%$, where $\rho_{xx}(0)$ is the in-plane resistivity at zero-field. (**B**) The tilt angle ($\theta$) dependent SdH oscillations, $\Delta\rho_{OSC}$. $\Delta\rho_{OSC}$ are extracted from the field dependent in-plane resistivity after a polynomial background subtraction. The top inset shows the derived SdH oscillation frequency as a function of the tile angle. The solid line is a fit assuming an ideal cylindrical Fermi surface geometry. The bottom inset is an illustration of the measurement geometry. The dashed line is normal to the plate-like sample plane.

## 2.7 Material parameters determined from QO measurements

Based on the results from the QO measurements, the size of each FS pocket can be estimated according to the Onsager relationship $F = (\Phi_0/2\pi^2)S_F$, where $\Phi_0$ is the magnetic flux quantum and $S_F$ is the cross-sectional area of the FS. The $\alpha$, $\beta_1$, $\beta_2$, $\gamma_1$ and $\gamma_2$ oscillations that were resolved in dHvA measurement correspond to 0.27%, 2.13%, 2.28% 3.67% and 3.82 % of the Brillouin zone (BZ) area, respectively, and the $\delta_1$ and $\delta_2$ oscillations that were additionally resolved from SdH measurement are 16.7 % and 17.8 % of BZ area, respectively. Furthermore, the cyclotron effective mass, Dingle temperature, and mobility that were extracted from the QO measurements are also shown in Table S2. The transport mobilities determined from Hall measurements (to be discussed below) are included for the convenience of comparison. Figure S8 shows the SdH oscillations measured on a 19-nm thin flake (Sample 7).

**Table S2**. Material properties derived from QO measurements. $F$, oscillation frequency; $m^*$, effective mass; $m_e$, free electron mass; $T_D$, Dingle temperatures; $\tau_q$, quantum lifetime; $\mu_q$, mobility derived from quantum lifetime.

| FS Pocket | $F$ (T) | $m^*/m_e$ | $T_D$ (K) | $\tau_q$ (×10$^{-14}$s) | $\mu_q$ (cm$^2$/Vs) | Method | Sample Geometry |
|---|---|---|---|---|---|---|---|
| $\alpha$ | 60 | NA | NA | NA | NA | dHvA | Bulk |
| $\beta_1$ | 472 | 0.174 | 7.2 | 16.9 | 1710 | | |
| $\beta_2$ | 506 | 0.162 | 9.0 | 13.5 | 1464 | | |
| $\gamma_1$ | 813 | NA | NA | NA | NA | | |
| $\gamma_2$ | 847 | 0.213 | 6.9 | 17.5 | 1446 | | |
| $\alpha$ | 61 | 0.106 | 18.9 | 6.4 | 1165 | SdH | Bulk* |
|  | (59) | NA | (10.0) | (12.1) | (2012) | | |
| $\beta_1$ | 473 | NA | NA | NA | NA | | |
| $\beta_2$ | 511 | NA | NA | NA | NA | | |
| $\gamma_2$ | 852 | NA | NA | NA | NA | | |
| $\delta_1$ | 3708 | NA | NA | NA | NA | | |
| $\delta_2$ | 3948 | NA | NA | NA | NA | | |
| $\alpha_1$ | 37 | NA | NA | NA | NA | SdH | Thin flake† |
| $\alpha_2$ | 49 | NA | NA | NA | NA | | |
| $\beta_1$ | 398 | NA | NA | NA | NA | | |
| $\beta_2$ | 421 | NA | NA | NA | NA | | |
| $\gamma_2$ | 839 | NA | NA | NA | NA | | |

*The material parameters for the $\alpha$ pocket are from Sample 3 and Sample 5 (in the parentheses, the same effective mass is assumed). The parameters for other pockets are from Sample 1.
†Material parameters from Sample 7.

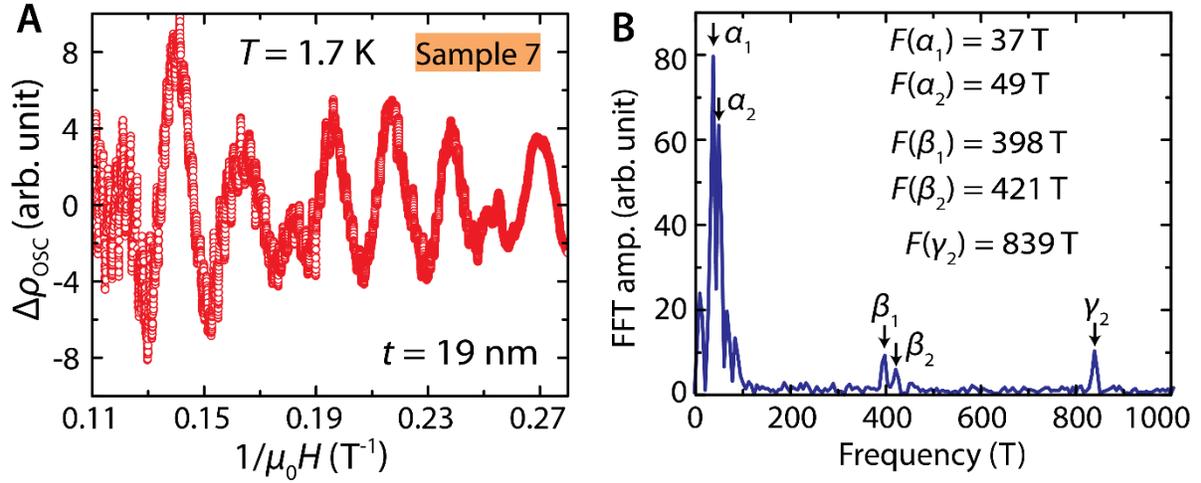

**Fig. S8. SdH oscillations measured on a 19-nm thin flake.** Note that two $\alpha$ pockets are identified and denoted as $\alpha_1$ and $\alpha_2$ in (B). The FFT spectrum of the SdH oscillation appears similar to that observed on a bulk sample (Fig. S6B). The $\delta_1$ pocket is not resolved in the thin flake either. Despite these common features, the FFT spectrum of the thin flake suggests an overall smaller size of all FS pockets (Table S2). Furthermore, two distinct $\alpha$ pockets are resolved and denoted as $\alpha_1$ and $\alpha_2$ in (B).

## 2.8 Carrier concentration and mobility by Hall measurements

Hall measurements were performed on both bulk samples and exfoliated thin flakes. Figure S9 shows the Hall resistivity ($\rho_{yx}$) and conductivity ($\sigma_{xy}$) with fittings from a two-band model. Multiple samples with different *RRR*s ranging from 67 to 825 were evaluated this way and their results are shown in Table 1 in the main text. The temperature dependent carrier concentration and mobility based on the fits to $\rho_{yx}$ on Sample 1 are plotted in Fig. 3, while those from fits to $\sigma_{xy}$ are plotted in Fig. S10. Results from both methods show a good agreement.

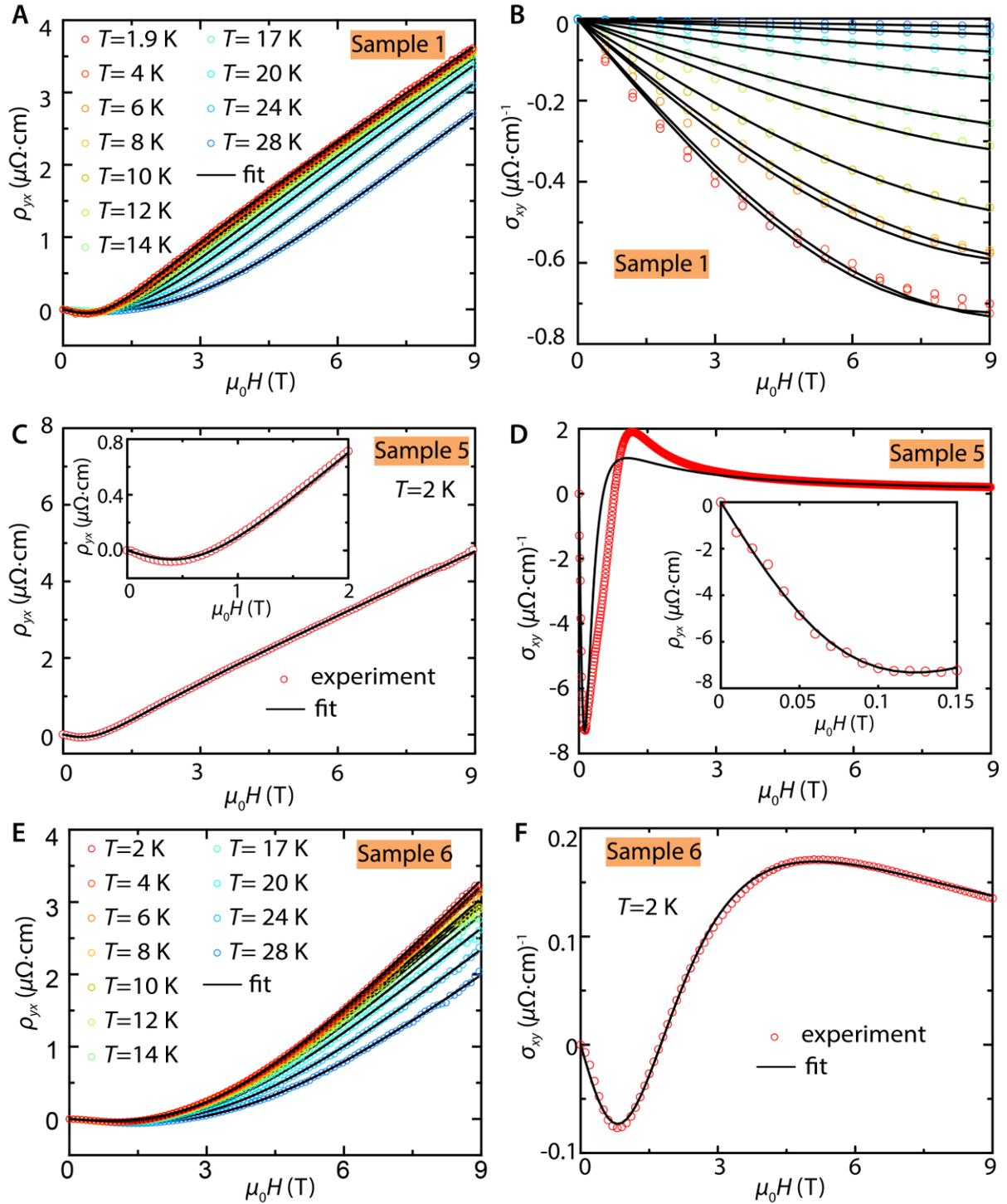

**Fig. S9. Two-band model fits to the Hall resistivity and conductivity at various temperatures on multiple samples.** (**A,B**) The Hall resistivity and low-field Hall conductivity,

respectively, and their fits for Sample 1 at various temperatures.. (**C,D**) The Hall resistivity and conductivity, respectively, measured on Sample 5 with two-band fits. The insets show the low-field region. The fit in (C) results in $n_e = 1.59 \times 10^{21}$ cm$^{-3}$, $n_h = 2.74 \times 10^{21}$ cm$^{-3}$, $\mu_t(e) = 113,000$ cm$^2$ V$^{-1}$ s$^{-1}$ and $\mu_t(h) = 15,000$ cm$^2$ V$^{-1}$ s$^{-1}$, while the fit in (D) results in $n_e = 2.28 \times 10^{21}$ cm$^{-3}$, $n_h = 3.43 \times 10^{21}$ cm$^{-3}$, $\mu_t(e) = 61,200$ cm$^2$ V$^{-1}$ s$^{-1}$ and $\mu_t(h) = 23,500$ cm$^2$ V$^{-1}$ s$^{-1}$. (**E,F**) Hall resistivity and conductivity, respectively, measured on Sample 6 with two-band fits.

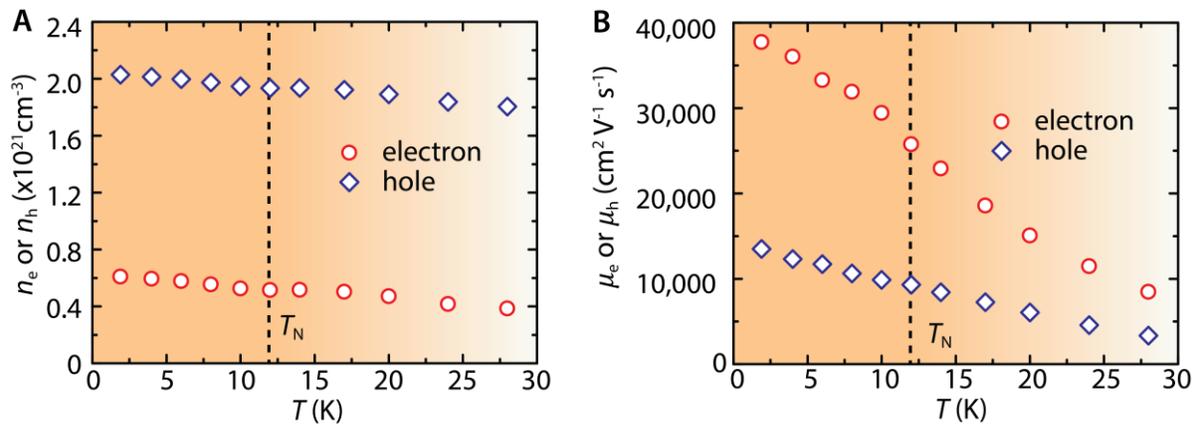

**Fig. S10. Temperature dependent carrier concentrations and mobilities from two-band model fits to the Hall conductivities measured on Sample 1.**

## 2.9 Comparison of the Fermi surface pockets from QO to calculated and measured electronic structures

Previously, band-structure calculations were performed for the unmodulated structure of LaTe$_3$ (*30*) and LuTe$_3$ (*62*) using the linear muffin-tin orbital (LMTO) method. These calculations revealed two electron-pockets encircling the X point of the BZ, two electron-pockets encircling Y and two hole-pockets encircling Γ and S (see Fig. S11 for a sketched version of the FS at the $k_z = 0$ plane). The doubling of each pocket arises from the bilayer splitting, similar to the well documented Fermi surface of Bi$_2$Sr$_2$Cu$_2$O$_{8+\delta}$ (*63*). If the CDW modulation is considered, the two pockets around X remain closed, as was suggested by the angle-resolved photoemission spectroscopy (ARPES) measurement on CeTe$_3$ (*64*). In the meanwhile, the FS pockets around Γ and S are partially gapped and those around Y are fully gapped by the CDW (Fig. S11, see also our ARPES data on GdTe$_3$ below). Particularly, band-structure calculations on LaTe$_3$ (*30*) predicted that the inner pocket around X has a size of 2.14% - 2.79% (depending on $k_z$) of the BZ area and the outer pocket around X has a size of 3.68% - 3.82% (depending on $k_z$) of the BZ area. These predictions match the size of the $\beta$ and $\gamma$ pockets observed in the dHvA and SdH oscillations well. Therefore, the pairing of $\beta_1/\beta_2$ pockets as well as $\gamma_1/\gamma_2$ pockets is attributed to the slight $k_z$ dispersions. Experimentally, the difference in $\beta_1$ and $\beta_2$ frequencies is smaller than that from calculations.

Based on the results from band-structure calculations, the hole-pockets are much larger than the electron-pockets of $\beta$ and $\gamma$. When the CDW modulation is considered, replica bands are expected to form in superposition to the original bands, and hybridizations of these bands can lead to the formation of new FS pockets, with a reduced size. This scenario was proposed for the explanation of the ARPES resolved FS in CeTe$_3$ (*39*), where the hole-pockets encircling Γ and S

are found to be significantly modified. Considering that $\gamma$ is the largest electron-pocket predicted in the calculations, we attribute the $\delta$ pocket to be a reconstructed hole-pocket. The small $\alpha$ pocket is also likely the product of FS reconstruction from the CDW.

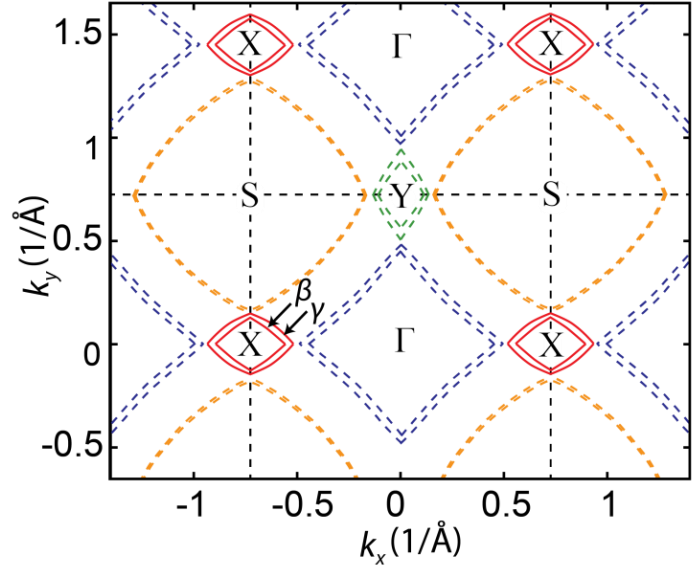

**Fig. S11. Schematic FS at the $k_z$ = 0 plane of bulk GdTe$_3$ based on the ARPES results and prior DFT calculations on LuTe$_3$.** Note that both pockets around X ($\beta$ and $\gamma$, solid lines) are not gapped by the CDW, while the pockets around Γ, S and Y (colored dashed lines) are partially or fully gapped by the CDW. The black dashed lines indicate the first BZ boundaries, not considering the CDW.

2.10 Carrier concentration estimations from QO measurements vs Hall measurements

The carrier concentrations can be estimated from the size of the FS pocket determined by QO measurements. Assuming a strict 2D geometry (no $k_z$ dispersion), the carrier density $n_{2D}$ per monolayer GdTe$_3$ per unit area is given by Luttinger's theorem as:

$$n_{2D} = 2\frac{S_F}{(2\pi)^2} = \frac{F}{\Phi_0}, \tag{S9}$$

where $S_F$ is the 2D cross-sectional area of FS, $\Phi_0$ is the magnetic flux quantum and $F$ is the QO frequency. Since we have detected a tiny $k_z$ dispersion, equation (S9) can be slightly modified to consider the $k_z$ dispersion as:

$$n_{2D} = \frac{S_{F1}+S_{F2}}{(2\pi)^2} = \frac{F_1+F_2}{2\Phi_0}, \tag{S10}$$

where $F_1$ and $F_2$ are the measured frequency pair. The 3D carrier concentration can be estimated taking the number of GdTe$_3$ layers per unit thickness into consideration. In the GdTe$_3$ structure, each monolayer is $c$ = 1.28 nm thick from our XRD measurement on the single crystals. Therefore, the 3D carrier concentration, $n_{3D}$, is calculated as:

$$n_{3D} = \frac{n_{2D}}{c} = \frac{F_1+F_2}{2\Phi_0 c} \tag{S11}$$

Using equation (S11), the overall electron carrier concentration is estimated to be $5.0 \times 10^{20}$ cm$^{-3}$ by considering the combination of $\beta$ and $\gamma$ pockets. The hole carrier concentration is estimated to be $1.5 \times 10^{21}$ cm$^{-3}$ by considering the $\delta$ pocket. Note that $\alpha$ is not considered for the carrier concentration estimation because of its negligible size. Overall, the carrier concentration values estimated from the QO frequencies are smaller than that determined from a two-band model fitting to the Hall data. This might be because the FS pockets are not fully resolved due to the low-field analysis performed here. We note that a previous optical conductivity measurement on GdTe$_3$ has suggested that only 3.2% of the original FS remains ungapped after the CDW modulation (*65*). Our current comprehensive QO and Hall measurements thus provide a more accurate description on the carrier concentration. The observation of the δ pocket on high-quality GdTe$_3$ crystals also provides an opportunity for better descriptions of the detailed FS geometry under the influence of the CDW, complementary to ARPES techniques.

2.11 Angle-resolved photoemission spectroscopy (ARPES)

Figure S12 shows the FS geometry at and slightly below the Fermi level, $E_F$. The Fermi velocity (Fig. S13) is estimated from two cross-sectional cuts (Path 1 and 2 in Fig. S12C)

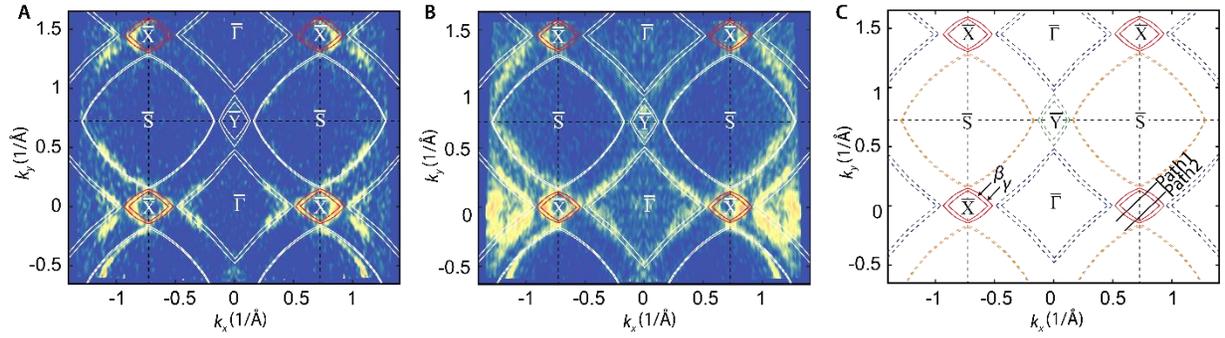

**Fig. S12. FS measured by ARPES and gap-opening by the CDW for GdTe$_3$.** (**A**) FS obtained by integration of the spectral weight in a 50 meV window around $E_F$. (**B**) The constant energy plots of the spectral weight in a larger window of 130 meV around $E = E_F$-0.18 eV. (**C**) Schematic FS of GdTe$_3$. The black dashed lines indicate the BZ boundaries of GdTe$_3$, not considering the CDW. The two solid black lines indicate the cross-sectional paths for the extraction of Fermi velocity, as shown in Fig. S13. Note: The constant energy plots are symmetrized with respect to $k_x$. The ARPES spectra were recorded with right circular polarized light at a photon energy of 500 eV.

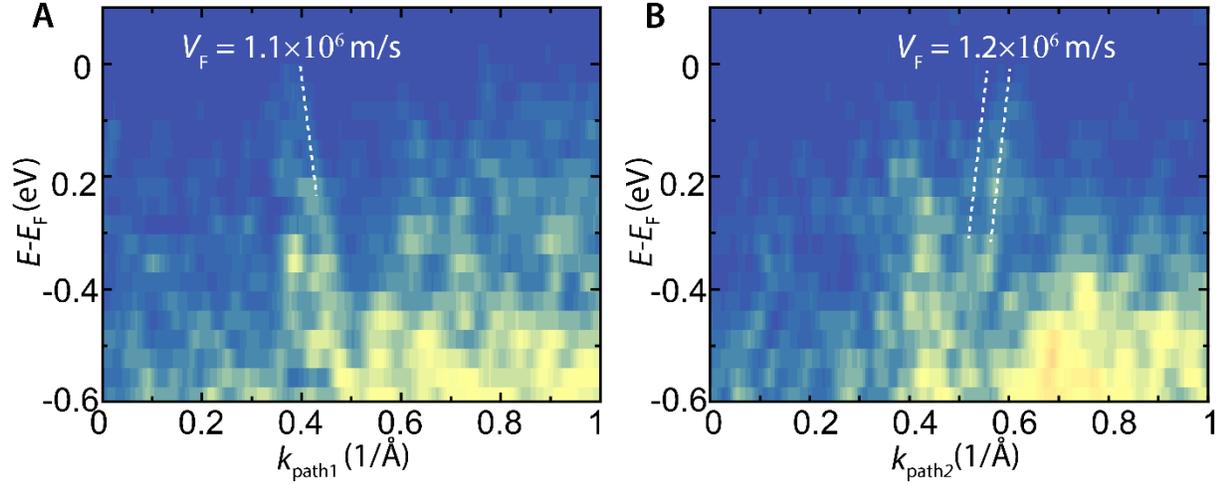

**Fig. S13. The Fermi velocity evaluations for the two X-pockets along X-Y directions.** (A,B) Band dispersion along Path 1 and 2, as indicated in Fig. S12C. The dashed lines represent linear fits to the bands near the Fermi level ($E_F$). The bilayer splitting is visible in **B**, as indicated by the two dashed lines. By assuming parabolic band dispersions, their effective masses are estimated to be: $m^*(\beta) = 0.11m_e \sim 0.13m_e$ and $m^*(\gamma) = 0.15m_e \sim 0.17m_e$, which is in close agreement with that determined from QO measurements.

2.12 Air sensitivity study and Raman spectroscopy of GdTe$_3$ thin flakes

The air sensitivity test was performed under ambient condition. While the thin flakes are stable in air for a short while, they start to degrade upon heating above ~100 °C or exposing in air for a longer time (>1 hours). Heating in inert atmosphere keeps the flakes intact, however (Fig. S14). The Raman spectroscopy of exfoliated GdTe$_3$ thin flakes was monitored in an inert atmosphere (Fig. S15). As expected for strongly absorbing samples, the Raman signal was strongly enhanced as the material was thinned. For thin flakes down to 15 nm the Raman signal is nearly un-altered, however for thinner samples a strong enhancement and small redshift was observed for a single mode at 125 cm$^{-1}$. Simultaneously, a mode near 120 cm$^{-1}$, seems to disappear. The origin for this behavior is currently not clear, but maybe the result of it merging with the intense mode. However, as all other modes remain close to their bulk values, we conclude the structure is intact. We note that the initial polarization and temperature dependent measurements on bulk suggest that the mode at 120 cm$^{-1}$ has the same symmetry and dependence on the CDW as the other modes, further suggesting the structure is largely un-altered by thinning.

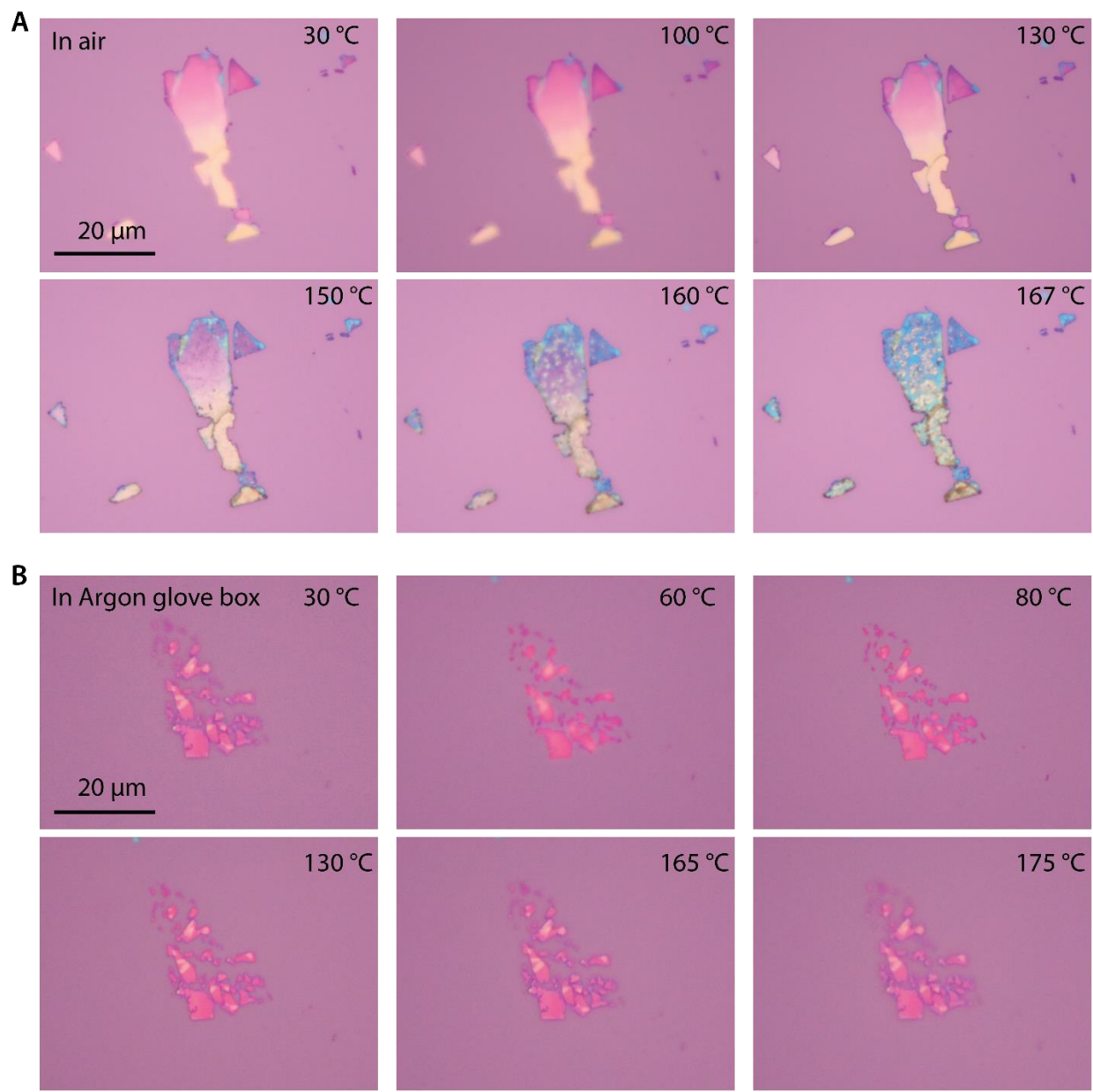

**Fig. S14. Air sensitivity of GdTe$_3$ thin flakes.** Heat treatment to above ~100 °C in air (**A**) will result in the degradation of thin flakes in a short while, but they remain stable when heated inside the argon-filled glovebox (**B**).

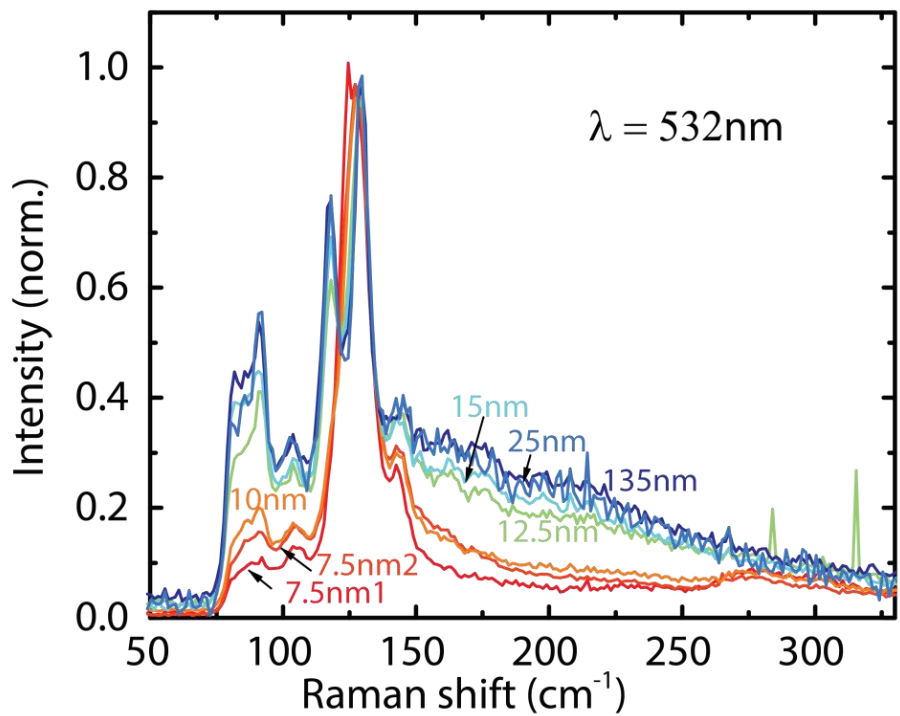

**Fig. S15. Raman spectroscopy on a series of GdTe$_3$ thin flakes with varying thickness.** No major change is observed in the Raman data with thickness. The Raman spectrum on 135 nm thick pieces resembles that of a bulk crystal.

## 2.13 Additional notes on mobility for materials shown in table 2

The mobility values that are listed in Table 2 can be classified into four categories: (1) mobilities ($\mu_q$) derived from quantum lifetime from QO measurements, (2) transport mobilities ($\mu_t$) determined from Hall measurement, (3) the transport mobilities from a combination of QO and residual resistivity measurements (denoted as "hybrid"), and (4) magnetoresistance. The mobility values determined from these methods can be different. The mobility derived from quantum lifetime is susceptible to both small- and large-angle scatterings, but the transport mobility is only sensitive to the large angle scattering (*36*). Therefore, the mobility measured from QO is generally smaller than the transport mobility. In the third method (the "hybrid method"), the QO frequencies are used to estimate the carrier concentration, which can be underestimated due to the possible unresolved FS pockets in QO measurements. With further one-carrier assumption ($1/\rho = ne\mu$), it leads to an overestimation of the averaged carrier mobility. The magnetoresistance method is sometimes used to estimate the mobility when the experimental data can be well characterized by a certain theoretical model.